\documentclass[osajnl,twocolumn,superscriptaddress,showpacs]{revtex4-1}
\usepackage{amsmath,amssymb,graphicx}

\begin{document}

\title{Surface plasmon polaritons scattering by  subwavelength silicon wires}

\author{Mehdi Shafiei Aporvari}
\email{Corresponding author:mshphy@gmail.com}
\affiliation{Department of Physics, Faculty of Science, University of Isfahan, Hezar Jerib, 81746-73441, Isfahan, Iran}

\author{Ahmad Shafiei Aporvari}
\affiliation{Department of Physics, University of Sistan and Baluchestan, Zahedan, 98135-674, Iran}

\author{Fardin Kheirandish}
\affiliation{Department of Physics, Faculty of Science, University of Isfahan, Hezar Jerib, 81746-73441, Isfahan, Iran}

\date{\today}
\begin{abstract}
Surface plasmon polaritons scattering from 2D subwavelength silicon wires is investigated using finite difference time domain method. It is shown that coupling an incident surface plasmon polariton to inter-cavity modes of the particle can dramatically changes transmitted fields and plasmon-induced forces. In particular, both transmission and optical forces are highly sensitive to the particle size that is related to the excitation of whispering gallery modes or standing-wave modes depending on the particle shape and size. This features might have potential sensing applications.
\end{abstract}
%\pacs{42.25.Bs, 87.80.Cc, 42.25.Ja, 42.79.Ag}
%\ocis{050.6624   Subwavelength structures; 290.0290   Scattering; 240.6680   Surface plasmons; 260.5740   Resonance}
\maketitle

 \section{Introduction}
Surface plasmon polaritons (SPP's) are evanescently confined modes that propagate at the interface between a dielectric and a conductor. This confinement leads to an enhancement of electromagnetic fields at the interface which is very sensitive to the surface defects \cite{polanco2013scattering, brucoli2011comparative}. The unique properties of the SPPs give them two dimensional nature and opens an opportunity for scaling down optical devices to nanometric dimensions \cite{maier2007plasmonics, gramotnev2010plasmonics}. In order to understand basic behaviors of such devices the interaction of SPPs with small particles and defects is a problem of fundamental importance. While, scattering by surface defects in all-metal structures has been studied in many works \cite{polanco2013scattering, brucoli2011comparative, pincemin1994scattering,nikitin2007scattering}, scattering by dielectric objects placed near a metallic surface has been explored less so far. The study of the latter structure might be particularly useful in the plasmonic optical tweezers in which a small object is trapped near a metallic surface \cite{min2013focused,zhang2014plasmonic, wang2009propulsion,righini2008light}. SPP scattering by finite-size gold nanocubes placed in the vicinity of a metal surface has been analyzed in Ref. \cite{evlyukhin2007surface}, and scattering by 2D rectangular dielectric particles has also been analyzed \cite{pincemin1994scattering}.
%Also, in Ref. \cite{pincemin1994scattering}, scattering by 2D rectangular dielectric particles has been studied.

While silicon is widely used  in conventional photonic devices, it also provides an important material for the design of silicon plasmonic devices with advanced functionalities. Resonant light interactions of high-refractive index dielectric nanoparticles have already been analyzed in details \cite{evlyukhin2014optical, van2013designing}.  Here, we consider their interactions with surface plasmon polaritons. It is well-known from Mie theory that for dielectric particles with higher refractive indexes, the light scattering efficiency increases. Due to these properties, high index dielectric material offer great practical advantages in many plasmonic structures such as low loss propagation and strong field confinement in the hybrid plasmonic waveguides \cite{oulton2008hybrid, yang2011optical, chen2012novel} and enhanced light transmission and enhanced trapping forces in the plasmonic nanoaperture traps \cite{garcia2002light, aporvari2015optical}. 

Dielectric microcavities can support whispering gallery modes (WGMs) which have been intensively applied to many devices  such as  optical sensors \cite{he2011detecting}, narrowband filters \cite{savchenkov2009narrowband}, and microlasers \cite{polman2004ultralow}.   Most research on  WGM cavities are restricted to large cavities with sizes much larger than light wavelength.  However, to develop advance nanophotonic elements, it is of great importance to consider nanosize resonators with sizes smaller than the considered light wavelength. In this regime there are few research on WGMs (see, for instance \cite{nobis2004whispering, cole2006easily}). In this paper we consider subwavelength silicon wires and investigate the excitation of WGMs with small mode numbers. 
%In this paper we consider subwavelenght silicon wireswhich due to their small sizes support low-order WGMs.

From the theoretical perspective, the scattering of surface plasmon polaritons by surface defects has been studied by several techniques such as the Green's Function method \cite{paulus2001light, pincemin1994scattering} and Rayleigh expansion method \cite{nikitin2007scattering,polanco2013scattering, zayats2005nano}. On the other hand, the finite difference time domain (FDTD) method is a powerful and well known technique 
to determine electromagnetic fields around nanostructures of complex and arbitrary geometries \cite{taflove2005computational,oubre2005finite, kuttge2009grooves, okamoto1999radiation}. In this paper,  we use two-dimensional FDTD method (a home-made code) for analyzing SPPs scattering from subwavelength silicon wires with different cross sections. We also use the Maxwell stress tensor (MST) technique to investigate the optical forces on the particles.  We show that SPP scattering parameters and optical forces on a subwavelength wire are non-trivial and strongly depend on its cross section  shape and size. Indeed, our calculations show that the excitation of intercavity modes such as WGMs and standing wave modes,  is responsible for this behavior. It is important to note that  the high refractive index plays a key role here and for sufficiently high refractive index material such as silicon, the subwavelength wire would exhibit WGMs (with small mode numbers) even if its size is smaller than the vacuum wavelength of the light.  Although they are less investigated, such low order modes in subwavelength regime could be of great importance for nanophotonics and nanosized devices \cite{nobis2004whispering}. 
\section{Calculation of Optical Forces}
To excite SPPs in the FDTD simulations we use the total-field/scatter-field technique \citep{taflove2005computational}. A small grid size (less than $3\,{\rm nm}$) is chosen in order to account for all plasmonic near-field behaviors. We bound the simulation domain with a convolutional perfectly matched layer (CPML) \cite{taflove2005computational}. For the dielectric constant of the metallic region, we use the Drude model with parameters that 
fitted to the experimental data of the complex dielectric constant for silver \cite{johnson1972optical}.
%dielectric data for silver \cite{johnson1972optical}. 

  Once the electromagnetic fields have been calculated using FDTD method, we can calculate the optical forces acting on the particle using the Maxwell stress tensor (MST) method \cite{novotny2012principles}. The time-averaged force acting on the center of mass of the particle is
\begin{equation}
\left<\textbf{F}\right>=\int_{S} \left<\mathbb{T}\right> \cdot \hat{\textbf{n}} \; dS
\end{equation}
where $S$ is a surface enclosing the particle, $\hat{\textbf{n}}$ is the unit vector perpendicular to the surface, and $\left<\mathbb{T}\right>$ is the time-averaged Maxwell stress tensor for harmonic fields, i.e.,
\begin{equation}
\left<\mathbb{T}\right>
=
\frac{1}{2} \Re 
\left\{
 \epsilon \textbf{E}\textbf{ E}^{*} +
 \mu \textbf{H}\textbf{ H}^{*} -
 \frac{\textbf{I}}{2} \left( \epsilon |\textbf{E}|^2 + \mu |\textbf{H}|^2 \right)
\right\} \; ,
\end{equation}
where $\textbf{E}$ is the electric field, $\textbf{H}$ is the magnetic field, and $\epsilon$ and $\mu$ are the permittivity and permeability of the surrounding medium.

We notice that for small particles or small entities in the structure (like small
gaps) and for low refractive index discrepancy between dielectric particle and surrounding, the
force calculations using the Maxwell stress tensor need a lot of care. In order to ensure the
correctness of our results we verified the numerical convergence of our algorithms. In our simulations we
considered a uniform two-dimensional square-cell space lattice with $\Delta x=\Delta z=\Delta$. For each configuration we performed a convergence analysis to ensure that the mesh size is
chosen properly (i.e. it is small enough) to reduce numerical errors. As a result of this
analysis we decided to employ a mesh size between 2 nm and 3 nm.

\section{Comparison of FDTD Computations and Rayleigh Expansion Approximation}
%In this section we consider scattering  of  SPPs by one-dimensional defects on a metal-vacuum interface. 
It is instructive to first compare FDTD computations with a theoretical model. We consider the scattering  of  SPPs by small defects on a metal-vacuum interface. 
As shown in Fig. 1 a SPP of frequency $\omega$ propagates along the \textit{x} axis. Suppose that the metal-vacuum interface is describe by the following function:
 \begin{figure}
\centerline{\includegraphics[width=0.9\columnwidth]{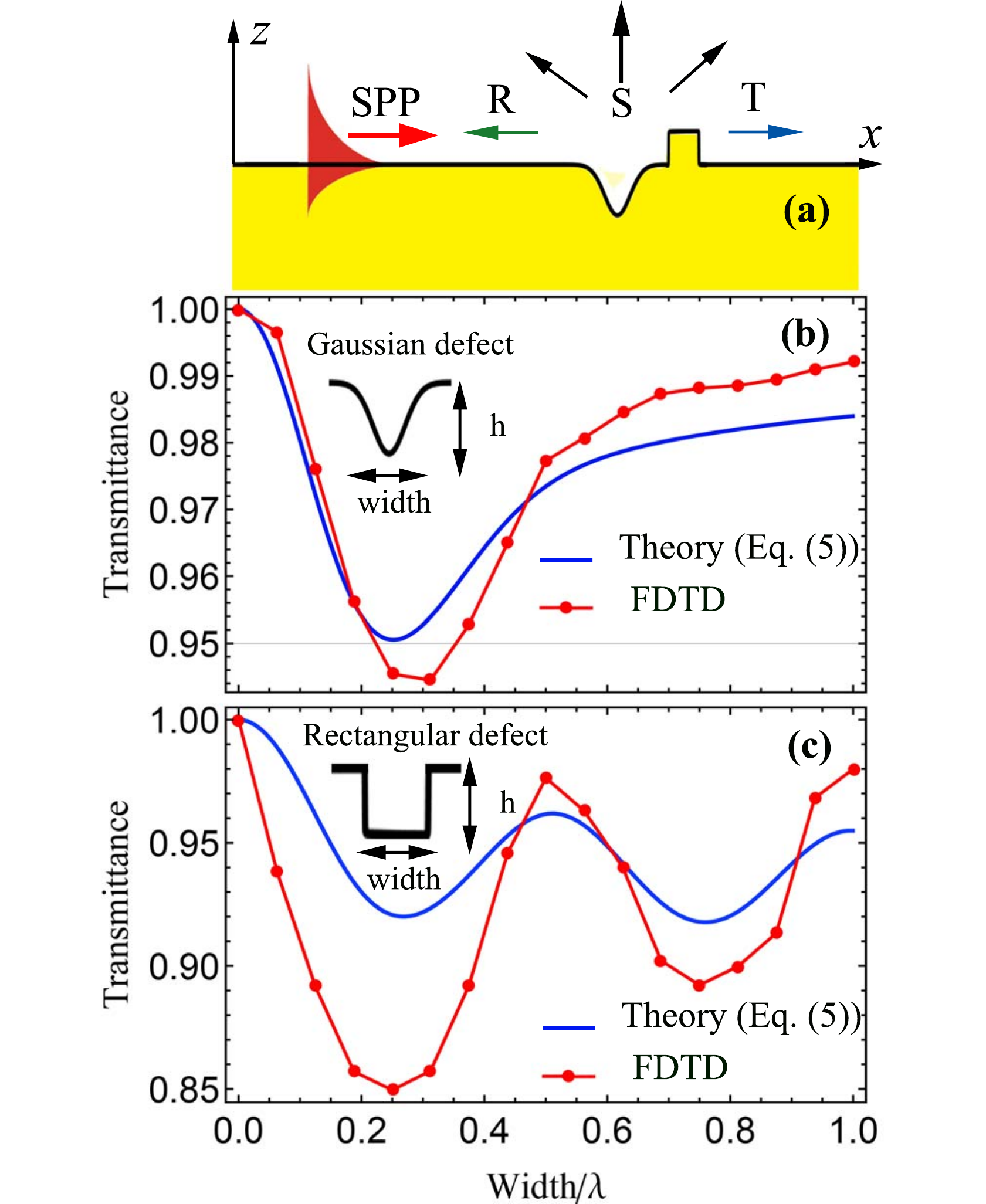}}
\caption{(a) Schematic of the surface plasmon scattering from surface defects. Transmittance as a function of defect width for  Gaussian (b) and rectangular (c) shape indentations in the silver-air surface ($\lambda=600~ \rm{nm}$ and $h_{max}=36 nm$). Solid curves correspond to Rayleigh approximation and circles represent the corresponding transmittance computed by FDTD method.}
\label{fig1}
\end{figure}

\begin{equation}
z=h(x).
\end{equation}
This form represents a one-dimensional defect in a single interface all-metal structure.
For small defects, the magnetic field in the vacuum half-space can be written as a sum of the incident field and scattered field according to the conventional Rayleigh expansion \cite{tsang2004scattering}:
\begin{equation}
H_y(x,z)=exp[ik_px-k_2z]+\int{dk H_y(k)exp[ikx-k_zz}]
\label{eq.rayleigh}
\end{equation}
where $k_p=k_0 \sqrt{\frac{\epsilon}{1+ \epsilon}}$, $k_2=k_0  \sqrt{\frac{ 1}{1+ \epsilon}}$ are the wave number of a surface plasmon polariton at a flat metal-vacuum interface and the inverse decay length of the fields perpendicular to the interface, respectively. Also,  $\epsilon$ is the dielectric function of the metallic half space, $k_0=\omega/c$, and $k_z$ is defined by $k_z=\sqrt{k^2-k_0^2}$, Re$\{k_z\}>0$.

In order to find scattered coefficients, $H_y(k)$, assuming that $|\epsilon|\gg 1$,  one can apply the surface impedance boundary condition (SIBC) to the tangential components of the fields at metal surface \cite{nikitin2007scattering}.
%\begin{equation}
%\textbf{E}_t(x,z)=\xi \textbf{H}_t(x,z) \times\textbf{ n}(x)
%\end{equation}
%where $\xi=\frac{1}{\sqrt{\epsilon}}$ is the surface impedance of metal and $\textbf{n}$ is the unite vector perpendicular to the surface. Considering the \textit{x} component of this equation yields
%\begin{equation}
%E_x(x,z)n_z(u) -E_z(x,z)n_x(u)=\frac{1}{\sqrt{\epsilon}}H_y(x,z), ~~\rm{at}~ z=h(x)
%\end{equation}
%Combining this and Maxwell equations $E_x=-i \frac{1}{\omega \epsilon_0\epsilon} \partial H_y / \partial z$   
%Rewriting Eq. \ref{eq.rayleigh} by making a change of variables $q=k/k_0$ and substitution of $r(q)$ and %applying Cauchy's integral theorem leads to the following reflected and transmitted SPP fields
As a result, the reflected and transmitted SPP fields may be written in the following form
\begin{gather} \label{transmission.eq}
H_y(x \rightarrow \infty, 0)=(1+\tau)exp(ik_px) \\
H_y(x \rightarrow -\infty, 0)=exp(ik_px)+\rho exp(-ik_px)
\end{gather}
%where
%\begin{equation}
%\tau =\frac{2\pi i \xi}{q_p},~~~ \rho=\frac{2\pi i \xi}{-q_p}.
%\end{equation}
%\textcolor{red}{where $\tau$ and $\rho$ are, respectively, the SPP transmission and reflection amplitudes.}  
where $\tau$ and $\rho$ are parameters to be calculated using the boundary conditions \cite{nikitin2007scattering}. 
Therefore, the transmission and reflection coefficients are
\begin{equation}
T=|1+\tau|^2, ~~~ R=|\rho|^2
\end{equation}
Also, the energy conservation law leads to scattering coefficient from $S=1-T-R$.

%Now, we can numerically solve Eq. \ref{integral_eq} to obtain reflection and transmission of a SPPs incident %on a surface defect. We also perform two-dimensional FDTD method to compare the results. 
% The results are shown in Fig. \ref{fig1} for two defect of rectangular (b) and Gaussian (c) profiles in a silver surface (at a wavelength of 600 nm). The FDTD results are in reasonable agreement with the theoretical prediction of Rayleigh approximation. However, for rectangular defect there is less agreement which could be attributed to the violation of the surface impedance boundary condition at the discontinuities points. 
Figure \ref{fig1} illustrates the results for two defect of rectangular (b) and Gaussian (c) profiles in a silver surface (at a wavelength of 600 nm). We also perform two-dimensional FDTD method to compare the results.
 As seen in the figure, there is a noticeable discrepancy between the FDTD results and the theoretical prediction of Rayleigh approximation. This discrepancy might be attributed to the simplified boundary conditions and ignoring the propagation losses in \eqref{transmission.eq}. Furthermore, here the defect depths , $h_{max}$, is relatively large. 
Note that, for rectangular defect there is much less agreement which could be attributed to the violation of the surface impedance boundary condition at the discontinuity points. However, for smooth and sufficiently shallow defects one would expect  the Rayleigh approximation leads to more accurate results \cite{nikitin2007scattering}.     

\begin{figure}
\centerline{\includegraphics[width=\linewidth]{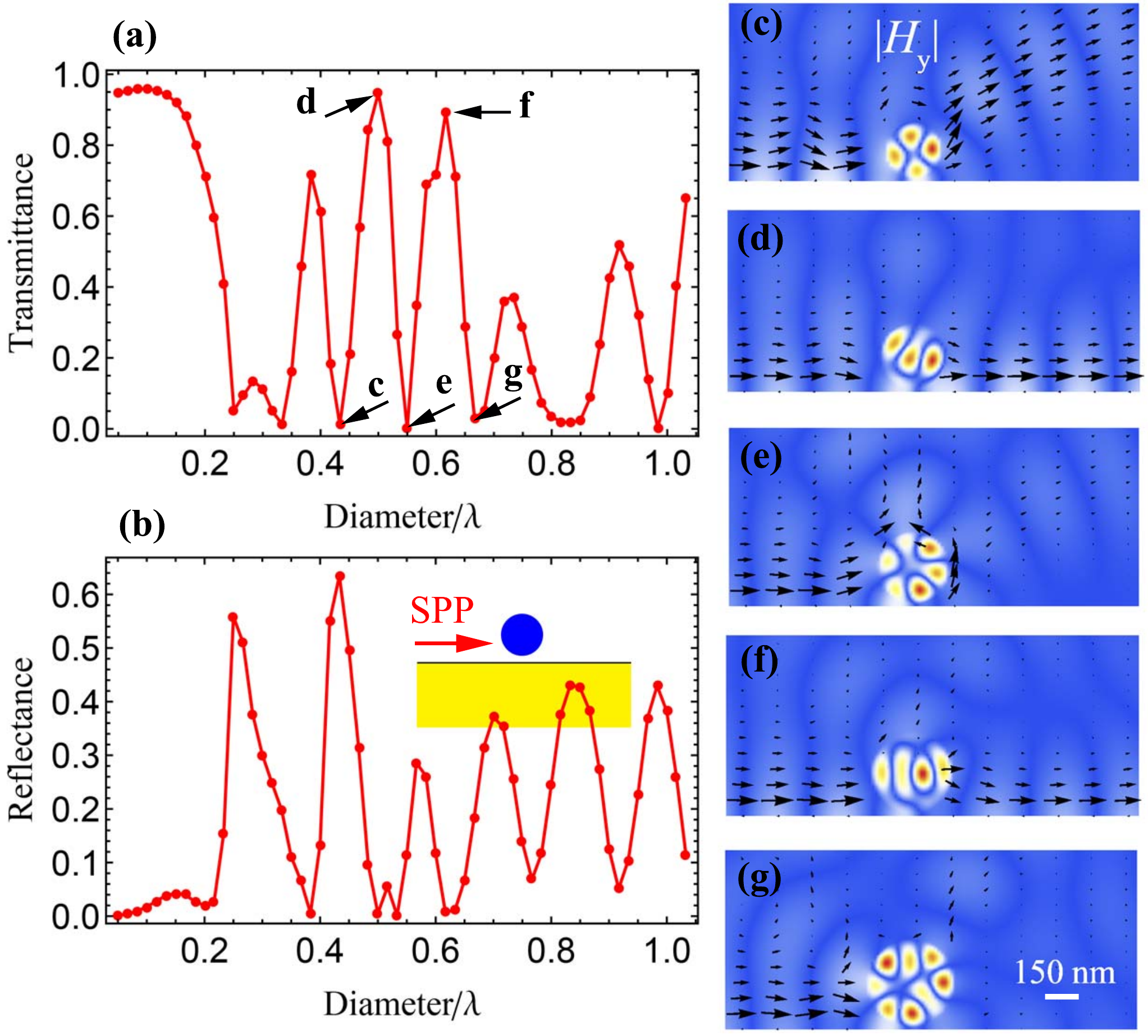}}
\caption{(a) Transmittance and (b) reflectance of a circular silicon cylinder as a function of its diameter near a flat silver surface at $\lambda=600$ nm. (c)-(g) near magnetic field patterns, $|H_y|$, in $z>0$ half-space for the particles are shown by arrows in Fig. 2 (a) . The arrows show the  Poynting vectors. The gap between the particle and silver surface is 12 nm.}
\label{fig2}
\end{figure}

\section{Scattering by subwavelength particles }
Let us now apply FDTD method to study the scattering of surface plasmon polaritons by  subwavelength silicon wires. 
We first consider scattering of a SPP from a circular silicon cylinder placed near a flat silver surface. The  surrounding   medium in the $z>0$ half-space is water (refractive index $n_m=1.33$), and the refractive index of Si was taken from Ref. \cite{palik1998handbook}. 
We fixed the gap between the particles and metal surface as 12 nm. Figs. \ref{fig2} (a) and (b) show transmittance and reflectance as a function of the particle diameter. As can be seen the particle behave in two different ways depending to its diameter. First, for very small particles ($D/\lambda <.15 $)   most incident SPP is transmitted, however larger particles transmit less power. Second, for larger particles ($D/\lambda >.15 $) the transmittance and reflectance curves strongly fluctuates. In this case, the transmittance suddenly rises and falls and surprisingly, in some points it is very high and in some points it is approximately zero.
 To better understand these behaviors, the magnetic field distribution ($|H_y|$) and Poynting vector field are shown in Figs. \ref{fig2} (c)-(g) for some particle of interest (depicted in Fig. \ref{fig2} (a) with arrows). For instance,  Fig. \ref{fig2} (c)  shows excitation of second-order (azimuthal) transverse magnetic (TM) polarized Whispering Gallery mode (WGM) by  evanescent coupling of incident SPP (here, the radius of the wire  is 130 nm, which corresponds to Diameter/wavelength=0.433).
More precisely, the whispering gallery modes, $\psi_{N,m} (r,\phi)$, of a 2D cylindrical cavity of radius $a$ can be analytically found by solving Helmholtz equation using the separation variable technique \cite{foreman2015whispering}. The solutions take the form,
\begin{equation}\label{wgm}
\psi_{N,m} (r,\phi)=A_{N,m} e^{\pm im\phi}G_m(k_{N,m} r),
\end{equation}
where the function $G_m$ is Bessel function of first kind, $J_m$, for $r<a$ and the Hankel function of the first kind, $H_m^{(1)}$, for $r>a$. Each mode is labeled as two mode numbers (N, m) which are the radial and azimuthal mode numbers. The mode numbers N and m are also related to the number of maxima in the field profile in the radial and azimuthal direction, respectively. Therefore, the field profile in Fig. \ref{fig2} (c) corresponds to the mode (1, 2). 
From field intensity and Poynting vector distribution, we can conclude that the excitation of resonant particle modes lead to a negligible SPP transmission. In this case, the coupling of incident SPP to the WGMs of the particle is very similar to the critical coupling in the microcavity-waveguide systems, in which the directly transmitted field and the field coupled out of resonator into the output waveguide are out of phase and thus vanish each other  \cite{vahala2004optical}. We remark that, similar plasmonic transmission behavior has been reported in the SPPs scattering by deep narrow grooves structured into flat gold surfaces \cite{kuttge2009grooves}, for which efficient coupling of incident SPPs to resonant cavity modes inside the grooves leads to high reflectivity.
  It is also clear from Poynting vector pattern in Fig. \ref{fig2} (c) that most incident power scatters to the radiation fields into the forward direction.  

In a simple geometric optics  picture, modes of order m correspond to a plane wave with propagation constant of $k=n k_0$, undergo multiple total internal reflection upon the interior surface of the resonator. After one full circulation within the cavity ($\phi=2\pi$ in \eqref{wgm}) we can write: 
\begin{equation}\label{ray_approx}
kL_m+m \phi_t=2\pi m,
\end{equation}
%$L_m=2ma \sin(\frac{\pi}{m})$
where $L_m=2ma \sin(\frac{\pi}{m})$, is the feedback path and $\phi_t$ is the polarization-dependent phase shift that occurs during each total internal reflection \cite{jackson1962classical}. Note that, the ray optics approximation is more suitable for large cavities, i. e. $a\gg \lambda$, however it can also describe low-order WGMs \cite{nobis2004whispering}.  For a fixed wavelength we see from \eqref{ray_approx} that $L_m \propto m$ which show that higher mode number needs longer paths and therefore larger cavities. 
  In fact, it is possible to excite higher order WGMs with increasing the particle diameter as demonstrated in Figs. \ref{fig2} (e) and (g) (for (N,m)=(1, 3) and (1, 4)). The Poynting vector field also demonstrate how the field scattered in different directions for different particles. Figures \ref{fig2} (d) and (f) illustrate the field patterns for nonresonant configurations where as can be seen most input power conveys to the transmitted surface plasmon polariton.  So, the coupling of SPP mode to inter cavity modes of the particle leads dramatically changes in the transmission, reflection and scattering coefficients.
This sensitivity of the transmitted SPPs to the particle size might be useful for single particle sensing down to subwavelength scales. In fact, the penetration of  the evanescent near-field to the upper dielectric medium makes the SPPs sensitive to the refractive index changes in their local environment. The presence of the particle can be easily determined because coupling between the SPP and particle causes a strong transmission change.  Here, all-optical excitation and detection is one of the most important advantages of SPP sensing. 
 
 We now investigate the optical force on the particles near a metal surface.
Large field enhancement near a metallic surface allows efficient optical trapping using evanescent plasmonic fields. Conventional optical traps have been widely used for manipulation of small dielectric particles and  living cells, although the size of trapped objects are usually larger than the light wavelength. Plasmonic optical traps make it possible to effectively trap smaller subwavelength particles on a flat surface at low input power \cite{aporvari2015optical}. Also, the surface plasmon radiation pressure can be used for  particle transport in a liquid environment \cite{wang2013theoretical}.  Here, we show how the particle sizes strongly affect the scattering and gradient plasmonic forces.
 
\begin{figure}
\centerline{\includegraphics[width=\linewidth]{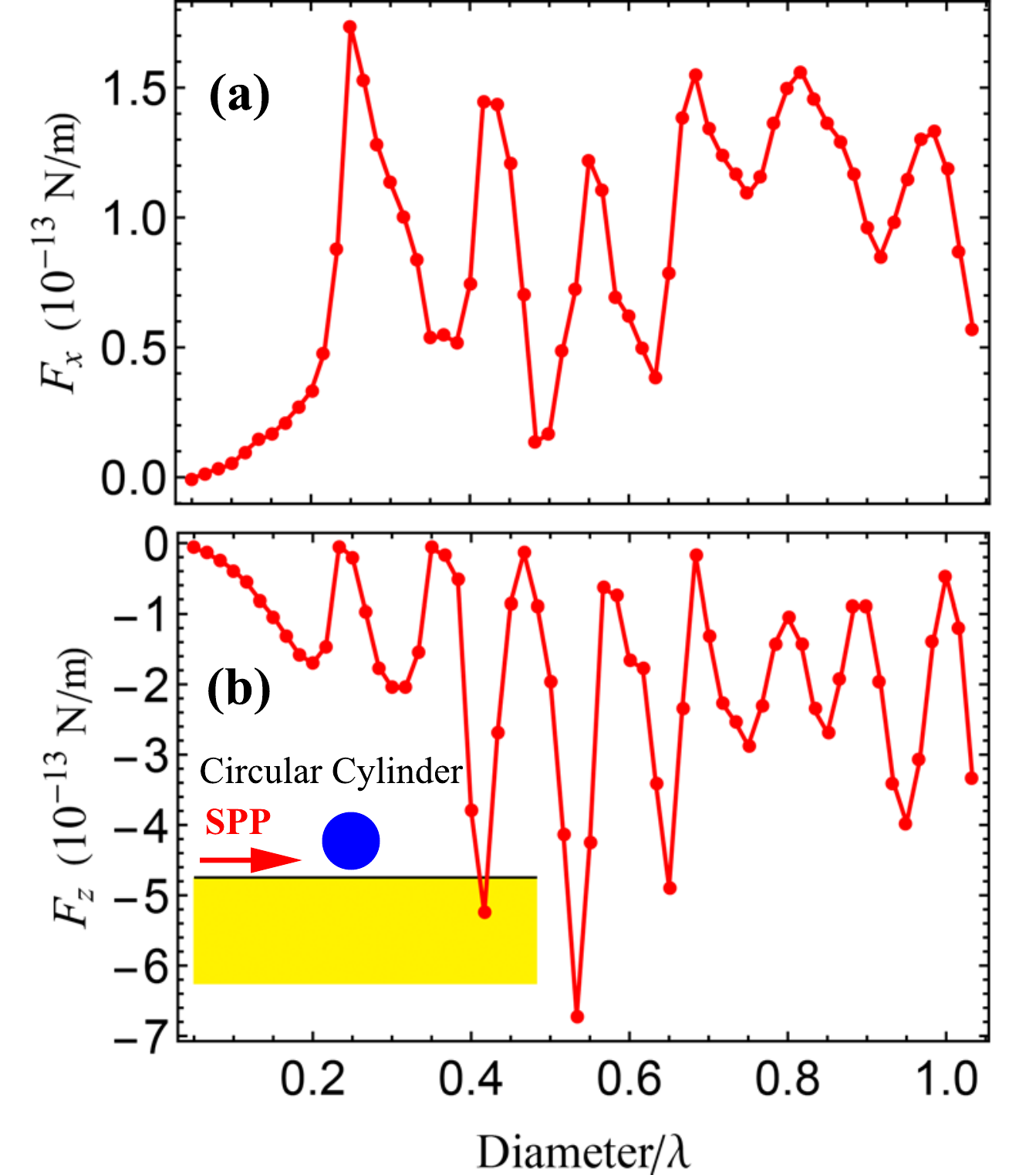}}
\caption{(a) Lateral optical force, $F_x$,  and (b) normal optical force, $F_z$, on a  circular silicon cylinder as a function of its diameter near a flat silver surface at $\lambda=600$ nm. The gap between the particle and silver surface is 12 nm.}
\label{fig3}
\end{figure}

 The normal and lateral optical forces has been demonstrated in Fig. \ref{fig3}. Again, we have two separate regains: small particle region where both $F_x$ and $F_z$ magnitudes are increased with increasing particle size and fluctuating force region. However, for very small particles, due to strong field gradient of the SPP wave perpendicular to the surface the attractive normal force, $F_z$ is much larger than scattered lateral force, $F_x$. In the fluctuating region we see the excitation of WG modes strongly affects the force magnitudes.  We should note that similar behavior was reported from an incident plane wave on spherical dielectric particles \cite{jia2009radiation}. However, for SPP incident wave the force fluctuations are much stronger compared to the plane wave incident.

\begin{figure}
\centerline{\includegraphics[width=\linewidth]{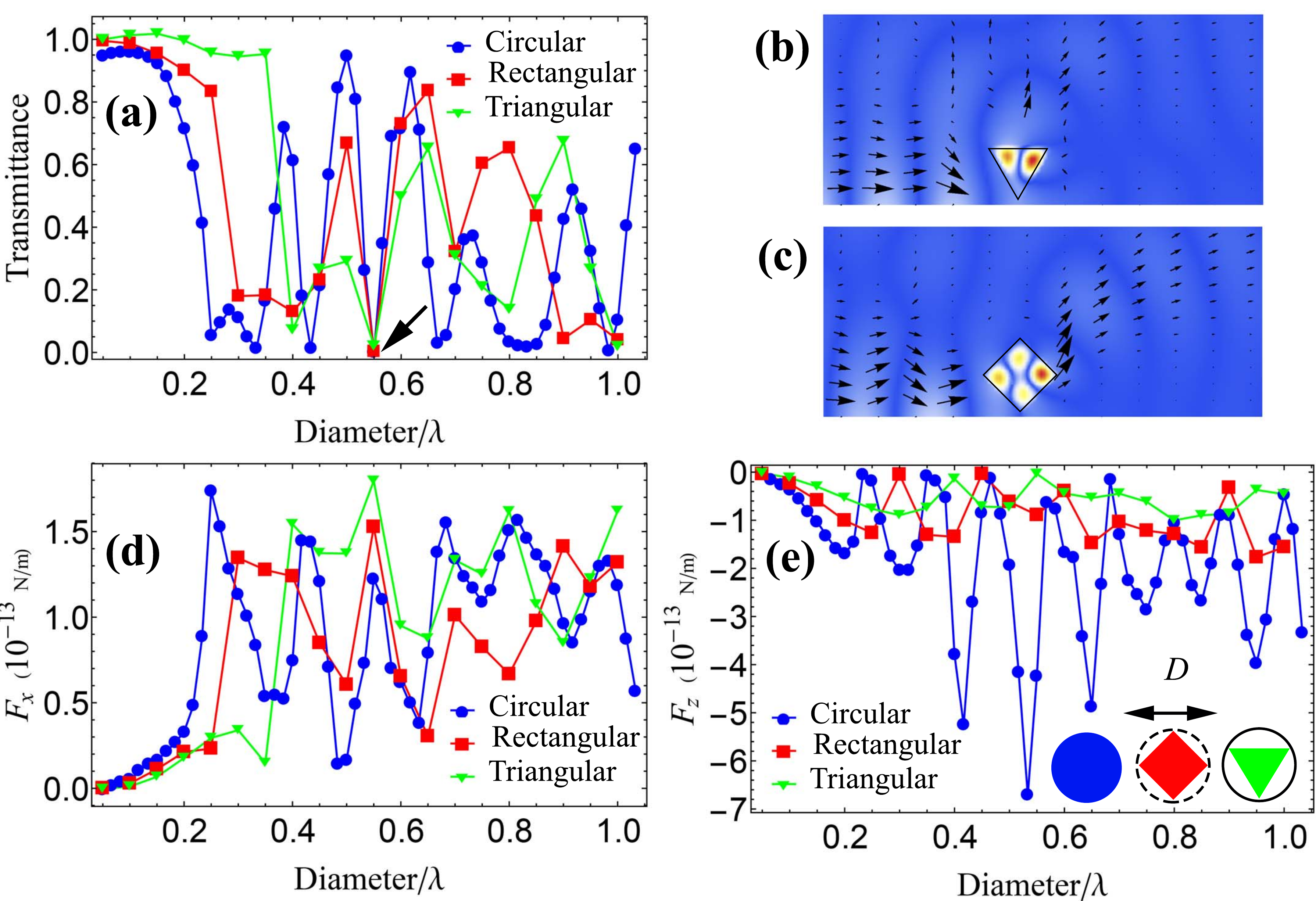}}
\caption{(a) Transmittance as a function of particle size, (b)-(c) magnetic field patterns, $|H_y|$ for the particles are shown by arrow in Fig. 4 (a), (d) the  optical forces $F_x$  and (e) $F_z$ for three different particle shapes near a silver-water surface ($\lambda=600~ \rm{nm}$ ). The gap between the particle (silicon) and silver surface is 12 nm. }
\label{fig4}
\end{figure}
Next, we shortly explore the influence of particle shape to the transmittance and optical forces. The results are shown in Fig. \ref{fig4}. For comparison we again show the circular particle transmittance and forces. The size of square and equilateral triangle particle are define as the diameter of the circle circumscribing the particle. As can be seen, the particle shape has considerable influence on the transmittance. However, notice that, for some particle sizes (see, e.g., the arrow in Fig. \ref{fig4} (a)), all three particles show a resonant behavior, but the excited modes are different. Their magnetic field patterns are illustrated in Fig. \ref{fig2} (e) and Figs.\ref{fig4} (b)-(c). It is possible to excite a standing wave mode for square particles. In fact, careful monitoring of the instantaneous magnetic field over one period reveals that the mode profile of the particle of Fig. \ref{fig4} (c) is a standing wave mode. The standing wave pattern can be given as $H_y(x',y')=A \sin(m_{x'}\pi x'/a) \sin(m_{y'}\pi y'/a) $, where $x'$ and $y'$ are coordinates in the  directions of the square sides of length a. We can thus deduce that Fig. \ref{fig4} (c) shows the excitation of  the mode $(m_{x'},~m_{y'})=(2,~2)$.

The induced optical forces are plotted in Figs. \ref{fig4} (d)-(e) which again show  high shape dependence.
Particularly, $F_z$ is much bigger for circular particles compared to the square and triangular particles.  

 \section{conclusions }
In conclusion, surface plasmons scattering by subwavelength defects and particles has been investigated using two dimensional finite difference time domain method. It is shown that even for small subwavelength particles the excitation of WG modes or standing wave modes strongly affects  the transmittance and reflectance of SPPs, which might have potential sensing applications. It is also shown that this mode excitation strongly influences the optical forces acting on the particle. While in this work we consider the 2D scattering problem, it can be straightforwardly extended to more general structures, e.g., SPPs scattering by nanorods at oblique incidence.  

% The authors wish to thank the graduate office of University of Isfahan for their support.

%\bigskip

% Bibliography
%\bibliography{sample}
%\bibliographystyle{unsrt}
%\bibliographystyle{apsrev4-1}
\bibliography{biblio}

%Manual citation list
%\begin{thebibliography}{1}
%\bibitem{Zhang:14}
%Y.~Zhang, S.~Qiao, L.~Sun, Q.~W. Shi, W.~Huang, %L.~Li, and Z.~Yang,
 % \enquote{Photoinduced active terahertz metamaterials with nanostructured
  %vanadium dioxide film deposited by sol-gel method,} Opt. Express \textbf{22},
  %11070--11078 (2014).
%\end{thebibliography}

\end{document}